\documentclass[prd,aps,preprint,epsfig,floats]{revtex4}

\newcommand{\beq}{\begin{equation}}
\newcommand{\eeq}{\end{equation}}
\newcommand{\beqn}{\begin{equation*}}
\newcommand{\eeqn}{\end{equation*}}
\newcommand{\bea}{\begin{eqnarray}}
\newcommand{\eea}{\end{eqnarray}}
\newcommand{\bean}{\begin{eqnarray*}}
\newcommand{\eean}{\end{eqnarray*}}
\newcommand{\ga}{\alpha}

\usepackage{graphicx}

\def\DESepsf(#1 width #2){\epsfxsize=#2 \epsfbox{#1}}

\begin{document}

\draft

\preprint{\vbox{
\hbox{UMD-PP-05-009}
}}

\title{{\Large\bf SO(10) Symmetry Breaking and Type II Seesaw }}
\author{\bf  H. S. Goh, R.N. Mohapatra and  S. Nasri }

\affiliation{ Department of Physics, University of Maryland, College Park,
MD-20742, USA}
\date{August, 2004}

\begin{abstract}
A minimal SO(10) model with {\bf 126} Higgs field breaking B-L
symmetry has been shown recently to predict large solar and
atmospheric mixings in agreement with observations if it is
assumed that the neutrino mass follows from the triplet dominated
type II seesaw formula. No additional symmetries need to be
assumed for this purpose. We discuss the conditions on the
way SO(10) symmetry breaks down to MSSM and the Higgs multiplets in the
model, required for the triplet dominated type II seesaw formula to
hold. We find that (i)
SO(10) must break to a nonminimal SU(5) before breaking to the
standard model; (ii) $B-L$ symmetry must break at the time of
SO(10) breaking and (iii)  constraints of unification
 seem to require that the
minimal model must have a {\bf 54} dimensional Higgs field
together with a {\bf 210} and {\bf 126} to break the GUT symmetry.
\end{abstract}

\maketitle

\vskip1.0in
\newpage

\section{Introduction}
The observed pattern of neutrino masses and mixings have posed a major
challenge for particle theory. While the seesaw mechanism
is emerging as a compelling way to
understand why neutrino masses are so small compared to charged lepton
and quark masses, there is no single accepted mechanism to
understand large mixings\cite{review}. In theories without quark-lepton
unification,
the large mixings may be a signal of new leptonic symmetries such as
 $L_e-L_\mu-L_\tau$ for inverted hierarchy or a
$\mu\leftrightarrow\tau$ discrete symmetry for normal hierarchy.
However, these symmetries cannot be imposed on theories that unify
quarks and leptons since there is no trace of them in quark masses
and mixings. Therefore understanding large mixing angles in the
context of grand unified becomes specially acute. Since grand
unified theories (GUT) have a number of interesting features
including the fact that the seesaw scale is very close to the GUT
scale, it is important to explore ways to understand the large
neutrino mixings in the GUT theories.

For neutrino masses, the most interesting grand unification models are
those based on the gauge group SO(10)\cite{so10} since (i) the {\bf 16}
dimensional spinor representation of the group that fits in matter of one
generation also contains the right handed neutrino that is an essential
part of the seesaw mechanism; (ii)
 the seesaw mass scale for the right handed neutrino which is determined
by the atmospheric neutrino data to be close to the GUT scale receives a
natural explanation as the GUT symmetry breaking scale and
(iii) SO(10) contains the Pati-Salam subgroup which
helps to connect the quark and lepton parameters thereby making the theory
potentially more predictive.

While these make the SO(10) models appealing for neutrino mass
studies, detailed quantitative predictions often require extra
symmetry assumptions beyond SO(10).  One exception to this is the
class of models that use only one {\bf 10} and one {\bf 126} Higgs
multiplet to generate fermion masses\cite{babu,goran,goh}. In this
case, if we ignore CP phases\cite{mimura}, there  are only 11
parameters describing the charged fermion masses and mixings.
Furthermore, the right handed neutrino mass matrix which
ordinarily depends on a new set of parameters, is now given by a
subset of the above 11 parameters plus an overall scale. This is a
sizable reduction in the number of parameters compared to the
standard model extended to include the seesaw mechanism, where we
have 31 parameters (30 in the absence of CP violation).

 If we further assume that in the type II seesaw
formula\cite{seesaw2} for neutrino masses that appears naturally in these
models, the triplet term dominates,
then the solar mass difference square and the two large mixing angles
$\theta_{12},
\theta_{23}$ are predicted and found to be consistent with present
observations at 2.5 to 3 $\sigma$ level and $\theta_{13}$ is
predicted to be $0.18$ which is slightly below the present
CHOOZ-Palo-Verde upper limit. The last prediction makes the model
testable in the next generation neutrino experiments.

 Crucial to the success of the model is the assumption that the triplet
term in the type II seesaw formula dominates over the second term.
In this paper we discuss the conditions under which this happens. We find
that they impose nontrivial constraints on the way SO(10) symmetry breaks
down to the minimal supersymmetric standard model (MSSM). In particular,
we find that SO(10) must break to the
standard model via a nonminimal SU(5) model. Secondly, the minimal
SO(10) model with the Higgs structure {\bf 10}, {\bf 126}
$\oplus{\overline{\bf 126}}$ and a {\bf
210} \cite{minimal,recent} needs to be extended by the addition of a {\bf
54} multiplet.

This paper is divided as follows: in sec. 2, we show that if
 the triplet term in the type II seesaw mass formula for neutrinos
is to dominate, the simplest way is to let SO(10) break to the standard
model via a nonminimal SU(5); in sec. 3, we discuss SO(10) breaking via
nonminimal SU(5) and discuss the spectrum of SU(5) representations; in
sec. 4, we point out that the model requires the the Higgs system to
contain a {\bf 54} dimensional field in addition to the {\bf 210},
{\bf 10} and a {\bf 126} pair to allow a light {\bf 15}-Higgs field;
 in sec. 5, we provide a detailed analysis of
symmetry breaking in the SO(10) model with {\bf 54}; in sec. 6, we
consider gauge coupling
unification and determine the values of the SU(5) and the SO(10) scales;
sec. 7 gives the conclusions of our study.

\section{Type II seesaw formula and SO(10) symmetry breaking}
It has been emphasized\cite{seesaw2} that in theories that conserve parity
asymptotically e.g left-right symmetric and SO(10) models, the seesaw
formula takes the form:
\begin{eqnarray}
{\cal M}_\nu \simeq f\frac{v^2_{wk}}{M_T}-\frac{m^2_D}{fv_{B-L}}.
\label{seesaw2}\end{eqnarray}
where $M_T\simeq \lambda v_{B-L}$, the mass of the triplet Higgs field and
the second term represents the formula for models that do not
have parity symmetry. This is called the type II seesaw formula. Note that
both the terms are inversely proportional to the scale $v_{B-L}$.
If in the above formula, the first term is assumed to dominate, then in
the minimal SO(10) models with only one {\bf 126}-Higgs (and arbitrary
number of {\bf 10}'s), one obtains a sumrule of the form\cite{brahma}
\begin{eqnarray}
{\cal M}_\nu~=~c(M_d-M_\ell)
\end{eqnarray}
In the context of minimal SO(10) models with only one {\bf 10} and
one {\bf 126}, this formula provides a very novel mechanism to
understand large neutrino mixings as was first pointed out in
Ref.\cite{goran} for the case of
 second and third generation neutrinos. The reason for large mixings is
 the known convergence of the $b$ and $\tau$ masses at the GUT
scale which for the case second and third generations leads to large
atmospheric neutrino mixing
angles  without further assumptions\cite{goran}. It was not clear at this
stage, whether the same mechanism works for the realistic three generation
SO(10) models. This was subsequently shown
in Ref.\cite{goh} that indeed the same $b-\tau$ mass convergence can also
explain large solar mixing angle and give a small $\theta_{13}$. 

As noted before, all these interesting results of minimal SO(10) are
two
assumptions : (i) breaking of B-L symmetry by a single {\bf 126} and
(ii) dominance of the triplet induced term in the type II seesaw for
neutrino masses. While qualitative arguments have been given in favor of 
the second assumption, a detailed investigation of this has not been
presented to date. In
this paper we fill this gap and critically analyze the second assumption
using the full Higgs structure of the model. 

To see that this is a nontrivial question, first note that which
term in Eq.\ref{seesaw2} dominates depends on the value of
$f_{ij}$ (since the two terms depend on it in different ways).
 We therefore need the value of
the Yukawa couplings $f_{ij}$. Let us first introduce the Yukawa
couplings of the model. If we denote the {\bf 10} Higgs field by
$H$, {\bf 210} by $\Phi$, $\overline{\bf 126}$({\bf 126}) Higgs
field by $\overline{\Sigma}$($\Sigma$) and the {\bf 16} spinor by
$\psi$, then the matter part of the superpotential consists only
of two terms:
\begin{eqnarray}
W~=~h_{ij}\psi_i\psi_j H +f_{ij} \psi_i\psi_j\overline{\Sigma}
\end{eqnarray}
At the GUT scale, this model has eight Higgs doublets (four up
type and four down type). By an appropriate doublet-triplet
splitting mechanism these four pairs are assumed to reduce to a
single MSSM Higgs pair $(\phi_u,\phi_d)$:
\begin{eqnarray}
\phi_u &=& \alpha_1^u H_{5} +  \alpha_2^u \overline{\Sigma}_{5}+
\alpha_3^u\Phi_{5} +
\alpha_4^u \Sigma_{45} \\
\nonumber \phi_d &=&  \alpha_1^d H_{\bar{5}} +  \alpha_2^d
\overline{\Sigma}_{\overline{45}}+ \alpha_3^d\Phi_{\bar{5}}
+\alpha_4^d \Sigma_{\bar{5}}
\end{eqnarray}
with the unitary condition $\sum_i|\alpha_i^{u,d}|^2=1$. As in the
case of MSSM, we will assume that the Higgs doublets $\phi_{u,d}$
have the vevs $<\phi^0_u>=v \sin\beta$ and $<\phi^0_d>=v
\cos\beta$, which then leads us to the mass formulae for quarks
and leptons.
\begin{eqnarray}
M_u &=& \bar{h} + \bar{f} \\ \nonumber
M_d &=& \bar{h}r_1 + \bar{f}r_2 \\ \nonumber
M_e &=& \bar{h}r_1 -3r_2 \bar{f} \\ \nonumber
M_{\nu^D} &=& \bar{h} -3 \bar{f}
\label{sumrule}\end{eqnarray}
where
\begin{eqnarray}
\bar{h} &=& 2 hv\overline{\alpha}_1^u \sin\beta \\ \nonumber
\bar{f} &=& \frac{1}{\sqrt{6}}f v\overline{\alpha}_2^u\sin\beta\\
\nonumber
r_1&=&\frac{\overline{\alpha}_1^d}{\overline{\alpha}_1^u}\cot\beta\\\nonumber
r_2 &=&
-\frac{2\overline{\alpha}_2^d}{\sqrt{3}\overline{\alpha}_2^u}\cot\beta
\end{eqnarray}
As a typical order of magnitude of the couplings $\bar{f}$,
$\bar{h}$, we consider the work in Ref.\cite{goh} and choose the
values of $r_{1,2}$ determined by the quark masses and mixings and
the masses of the charged leptons and find for our choice of
parameters\cite{goh}
\begin{eqnarray}
h~=~\pmatrix{3.26\times 10^{-6} & 1.50\times 10^{-4} & 5.51\times
10^{-3} \cr 1.50\times 10^{-4} & -2.40\times 10^{-4} & -0.0178 \cr
5.51\times 10^{-3} & -0.0178 & 0.473}
\end{eqnarray}
and
\begin{eqnarray}
f~=~\pmatrix{-7.04\times 10^{-5} & -2.05\times 10^{-5} & -7.53
\times 10^{-4} \cr  -2.05\times 10^{-5} & -1.85\times 10^{-3} &
2.43\times 10^{-3} \cr -7.53 \times 10^{-4} & 2.43\times 10^{-3} &
-1.64\times 10^{-3} }.
\end{eqnarray}
To get an idea of what kind of requirements are imposed by type II
seesaw, note that the largest element in the matrix $f$ is $\sim
\times 10^{-3}$ whereas as that in $h$ is about $0.5$. From this
we estimate that the biggest contribution to neutrino mass from
the second term (the canonical seesaw term) is about $\sim
\frac{m^2_t(M_U)}{f_{ij,max}v_{B-L}}\simeq \frac{3\times
10^{6}}{v_{B-L}}$ GeV. For $v_{B-L}\simeq M_U\simeq 2\times
10^{16}$ GeV, this gives an estimate for $\sqrt{\Delta
m^2_{A}}\simeq 0.15$ eV which is slightly bigger than the
experimental value. If type II seesaw is to hold this must be much
smaller than $ \sqrt{\Delta m^2_A}\simeq 0.05$ eV. If we take this
number to be 0.02 eV, it would require a value of $v_{B-L}\simeq
10^{17}$ GeV. Furthermore, for the first term to give the correct
value for $\sqrt{\Delta m^2_A}$, the mass of of the color singlet,
$SU(2)_L$ triplet Higgs field (denoted henceforth by $\Delta_L$)
should be of order $10^{12}$ GeV. Clearly the presence of such a
light triplet is going to affect unification of couplings.

The above remarks have the following implications:

\begin{itemize}

\item First and foremost is that $v_{B-L} \geq 10^{17}$ GeV which can
happen if $SO(10)$ first breaks
to SU(5) which subsequently breaks to the standard model.\\

\item Second is that maintaining gauge coupling unification would dictate
that the $SU(2)_L$ triplet $\Delta_L$ be part of a complete SU(5) {\bf 15}
multiplet at the same scale.

\end{itemize}

We will explore under what conditions, these requirements are
satisfied in our SO(10) model. We must note that in deriving the
above estimates for $v_{B-L}$ and $M_{\Delta_L}$, we have used the
numerical values for the Yukawa couplings from the neutrino fit in
\cite{goh} Eq.(1). However, one could consider variations of type
II seesaw SO(10) models that contain a {\bf 120}
multiplet\cite{mimura1} where the value of $f_{33}$ is of order
$0.1$. In such a case, even a $v_{B-L}\simeq 2\times 10^{16}$ GeV
can lead to a triplet dominated type II seesaw, provided
$M_{\Delta_L}\simeq 2\times 10^{14}$ GeV. In this case also one needs both
these conditions to be satisfied though at a somewhat milder level.


 We now proceed to discuss how these conditions can be satisfied in the
SO(10) model that first breaks to nonminimal SU(5) and then to the
standard model.

\section{Breaking SO(10) to standard model via SU(5)}

We start with the minimal Higgs fields  $H$({\bf 10}), $\Phi$({\bf
210}), $\Sigma$({\bf 126}) and $\overline{\Sigma}$($\overline{\bf
126}$) and write down the most general renormalizable
super-potential: \beq
    W~=~\frac{m_{\Phi}}{2\times 4!}\Phi^2 +
    \frac{m_{\Sigma}}{5!}\Sigma\overline{\Sigma}+\frac{m_H}{2} H^2 \\
     +\frac{\lambda}{4!}\Phi^3+
\frac{\eta}{4!}\Phi\Sigma\overline{\Sigma}+\frac{1}{4!}\Phi
     H(\ga \Sigma+\overline{\ga}\overline{\Sigma})
\eeq
We then extract the various SU(5) submultiplets from each of the
SO(10) Higgs multiplets and rewrite the superpotential in terms of these
fields. Extensive discussion of the decomposition of SO(10) multiplets in
terms of its subgroups \cite{nath,goran1,aulakh,fukuyama} as well as
detailed analysis of the potential exists in the
literature\cite{goran1,aulakh,fukuyama}. We have calculated the SU(5)
decomposition of various SO(10) invariant couplings and use them in this
paper. For this purpose note that
\begin{eqnarray}
{\bf 210}~=~ {\bf 1}_0 \oplus {\bf 5}_{-8} \oplus \overline{\bf
5}_{+8}\oplus {\bf 10}_4 \oplus \overline{\bf 10}_{-4} \oplus {\bf
24}_0 \oplus {\bf 75}_0
\oplus {\bf 40}_{-4}\oplus \overline{\bf 40}_{+4}\\
{\bf 126}~=~{\bf 1}_{-10}\oplus {\bf \overline{5}}_{-2} \oplus
{\bf 10}_{-6}\oplus {\bf \overline{15}}_{+6}\oplus {\bf 45}_2
\oplus {\bf \overline{50}}_{-2}
\end{eqnarray}
In terms of the properly normalized SU(5) submultiplets we now
rewrite first the bilinear terms and then the trilinear terms in
the superpotential 
 \bea
    L_B&=&m_H H_aH^a
    +m_{\Phi}\{(\Phi^5)_a(\Phi^5)^a+
\frac{1}{3!}{(\Phi^{40})_{abc}}^d{(\Phi^{40})^{abc}}_d\\\nonumber
    &+&\frac{1}{2}(\Phi^{10})_{ab}(\Phi^{10})^{ab}+
\frac{1}{8}{(\Phi^{75})_{ab}}^{cd}{(\Phi^{75})_{cd}}^{ab}
    +\frac{1}{2}{(\Phi^{24})_{a}}^{b}{(\Phi^{24})_{b}}^{a}+\frac{1}{2}
    \Phi_o^2\}\\\nonumber
    &+&m_{\Sigma}\{\sigma_o\overline{\sigma}_o+
\frac{1}{4!}{(\overline{\Sigma}^{15})_{a}}^{bcde}{(\Sigma^{15})^{a}}_{bcde}
+\frac{1}{2!}(\overline{\Sigma}^{10})_{ab}(\Sigma^{10})^{ab}\\\nonumber
    &+&\frac{1}{12}{(\overline{\Sigma}^{50})_{abc}}^{de}
{(\Sigma^{50})^{abc}}_{de}
    +\frac{1}{2}{(\overline{\Sigma}^{45})_{a}}^{bc}{(\Sigma^{45})^{a}}_{bc}
+(\overline{\Sigma}^{5})_{a}(\Sigma^{5})^{a}\} \eea

and the trilinear term become

\bea
   L_T&=& \frac{\lambda}{\sqrt{10}} \Phi_o\{12 (\Phi^5)^a(\Phi^5)_a +
    3(\Phi^{10})_{ab}(\Phi^{10})^{ab}+{(\Phi^{24})_a}^b{(\Phi^{24})_b}^a-
\frac{1}{2}{(\Phi^{75})_{ab}}^{cd}{(\Phi^{75})_{cd}}^{ab}\}\\\nonumber
    &+&\overline{\ga}
\overline{\sigma}_o H^a(\Phi^5)_a+
\overline{\ga}\sqrt{\frac{3}{5}} \Phi_o H^a(\overline{\Sigma}^5)_a
+ \ga \sigma_o H_a(\Phi^5)^a+
    \ga\sqrt{\frac{3}{5}} \Phi_o H_a(\Sigma^5)^a
\\\nonumber
    &+&\sqrt{6}\eta \bar{\sigma}_o\{
    (\Phi^5)_a(\Sigma^5)^a+
\frac{1}{2}(\Phi^{10})^{ab}(\Sigma^{10})_{ab}\}
 +c.c.\\\nonumber
    &+&\frac{\eta}{\sqrt{10}} \Phi_o \{
    2(\Sigma^{10})^{ab}(\overline{\Sigma}^{10})_{ab}
    +\frac{1}{12}{(\Sigma^{15})^a}_{bcde}{(\overline{\Sigma}^{15})_a}^{bcde}
    -\frac{1}{3!}{(\Sigma^{50})^{abc}}_{de}
{(\overline{\Sigma}^{50})_{abc}}^{de}
+4(\Sigma^{5})^a(\overline{\Sigma}^{5})_a\}\\\nonumber
    &+&\frac{2\lambda}{\sqrt{10}} \Phi_o^3  +\eta\sqrt{10}
\Phi_o\sigma_o\overline{\sigma}_o \eea
\subsection{SYMMETRY BREAKING AND RELATIONS AMONG THE PARAMETERS}
We can now discuss SO(10) breaking to SU(5). There are three SU(5)
singlets: one in each of the {\bf 126} pair and one in {\bf 210}.
The SU(5) singlets in the {\bf 126} pair have nonzero B-L and
therefore B-L breaking scale is same as the SO(10) scale. Since
supersymmetry must remain unbroken all the way down to the weak
scale, we set the F-terms to zero. These F conditions give the
following constraints on the vacuum expectation values:
\bea
    F_{\Phi_o}=m_{\Phi}\widetilde{\Phi}_o+6\lambda
    \widetilde{\Phi}_o^2+\eta
    \sigma_o\overline{\sigma}_o=0\\\nonumber
    F_{\overline{\sigma}_o}=\sigma_o(m_{\Sigma}+10 \eta\widetilde{\Phi}_o)=0
\eea where $\widetilde{\Phi}_o=\frac{\Phi_o}{\sqrt{10}}$ . The
solution that breaks B-L is

\bea
    \widetilde{\Phi}_o&=&-\frac{m_{\Sigma}}{10\eta} \\\nonumber
     \sigma_o\overline{\sigma}_o&=&
\frac{m_{\Sigma}}{10\eta^2}(m_{\Phi}-\frac{3\lambda
m_{\Sigma}}{5\eta}) \eea Note that with this minimal set of Higgs
fields, $\sigma_o\overline{\sigma}_o$ has to be non-vanishing in
order to get the standard model group below the GUT scale because
the $\sigma_o$, $\overline{\sigma}_o$ are the only singlets that
break the local B-L.
\subsection{MASSES OF SU(5) SUB-MULTIPLETS}

From the Lagrangian found above we easily write down the masses of
the various SU(5) submultiplets; we list those with no mixing in
Table I.
\begin{center}
{\bf Table I}
\end{center}

\begin{center}
\begin{tabular}{|c|c|}
  \hline
  40 & $m_{\Phi}$ \\
  \hline
  75 & $m_{\Phi}-4\lambda \widetilde{\Phi}_o$ \\
   \hline
  24 & $m_{\Phi}+2 \lambda \widetilde{\Phi}_o$ \\
   \hline
  15 & $\frac{4}{5}m_{\Sigma}$ \\
   \hline
  50 & $\frac{6}{5}m_{\Sigma}$ \\
   \hline
  45 & $m_{\Sigma}$ \\
  \hline
\end{tabular}
\end{center}

\noindent{\bf Table caption:} This table gives the masses of the various
SU(5) multiplets in the SO(10) multiplets of the minimal model.

\bigskip

The mass matrix for the SU(5) singlets in the basis
$(\Phi_0,\sigma_0, \overline{\sigma}_0)$ is found to be: \beq
\left(%
\begin{array}{ccc}
  m_{\Phi}+12 \lambda \widetilde{\Phi}_o  & \sqrt{10} \eta \overline{\sigma}_o
& \sqrt{10} \eta \sigma_o  \\
  \sqrt{10} \eta \overline{\sigma}_o  & 0 & 0 \\
  \sqrt{10} \eta \sigma_o  & 0 & 0 \\
\end{array}%
\right) \eeq One of the combination of the singlets has zero mass
and is the Goldstone Boson corresponding to the breaking of B-L.
As can be seen from the above matrix, the corresponding field is a
linear combination of the fields $\overline{\sigma}_o$ and
$\sigma_o$. Taking this out, we find the $2\times 2$ mass matrix
to be: \beq
\left(%
\begin{array}{cc}
  m_{\Phi}+12 \lambda \widetilde{\Phi}_o  &  \sqrt{10} \eta
\sqrt{\overline{\sigma}_o^2+\sigma_o^2} \\
 \sqrt{10} \eta \sqrt{\overline{\sigma}_o^2+\sigma_o^2}   & 0  \\
\end{array}%
\right)
\eeq
The mass eigenvalues are given by:
\beq
    m_{singlet}~=~\frac{m_{\Phi}+ 12 \lambda \widetilde{\Phi}_o \pm
\sqrt{(m_{\Phi}+12 \lambda \widetilde{\Phi}_o)^2
    +40 \eta^2(\overline{\sigma}_o^2+\sigma_o^2)}}{2}
\eeq

The mass matrix for the $\textbf{10}$ is
\beq
\left(%
\begin{array}{cc}
  m_{\Phi}+6 \lambda \widetilde{\Phi}_o & \sqrt{6} \eta \overline{\sigma}_o \\
  \sqrt{6} \eta \sigma_o & -6\eta \widetilde{\Phi}_o \\
\end{array}%
\right) \eeq This mass matrix has a zero eigenvalue and the
associated eigenstate field is the Goldstone boson corresponding
to the breaking of SO(10) down to $SU(5)\times U(1)$. The massive
combination has mass \beq
    m_{10}~=~-\frac{\eta}{\widetilde{\Phi}_o}(|\overline{\sigma}_o\sigma_o| +6
\widetilde{\Phi}_o^2) \eeq 

The mass matrix for $5$-plet Higgs in
the basis of ${\bf 5}=(H,\Phi,\overline{\Sigma})$ and $\bar{\bf
5}=(H,\Phi,\Sigma)$ is

\beq\label{5mass}
\bar{\bf 5}\left(%
\begin{array}{ccc}
  m_H & \overline{\alpha}\overline{\sigma}_o & \sqrt{6}\overline{\alpha}\widetilde{\Phi}_o \\
  \alpha\sigma_o & m_{\Phi}+12 \lambda \widetilde{\Phi}_o & \sqrt{6}
\eta \sigma_o \\
  \sqrt{6}\alpha\widetilde{\Phi}_o & \sqrt{6} \eta \overline{\sigma}_o &
 -6\eta \widetilde{\Phi}_o \\
\end{array}%
\right){\bf 5} \eeq


\section{NECESSITY OF {\bf 54} HIGGS FIELD}
It is clear that in the minimal model with {\bf 10}, {\bf 126}
pair and a {\bf 210} the masses of the SU(5) submultiplets {\bf
15}, {\bf 50} and {\bf 45} are proportional to the same parameter
$m_\sigma$. Therefore if we want to have {\bf 15} Higgs fields at
the sub-SU(5) scale of $10^{13}$ GeV, in order to enforce the type
II seesaw formula with the triplet vev dominating, then we would
have to have also the {\bf 45} pair and the {\bf 50} pair at
nearly the same scale. This however will affect the evolution of
gauge couplings very drastically.
 We therefore need a way to split
only the {\bf 15} dim. field without affecting the other fields. As we
show below, this is precisely what happens if we add to the model an
additional {\bf 54} dimensional Higgs field. The main reason for this is
that the {\bf 54} Higgs field contains an additional SU(5) {\bf 15} Higgs
field.

In the presence of the {\bf 54} Higgs field (denoted by $S$), the
superpotential of Eq. $(9)$ has the following additional terms:

\bea
    W_{54}&=&\frac{m_{15}}{4}S_{ab}S_{ab}+\frac{\lambda_1}{3!}
    S_{ab}S_{bc}S_{ca}+\frac{\lambda_2}{2}S_{ab}H_aH_b\\\nonumber
    &+&\frac{\lambda_3}{2\times 4!}
    S_{ab}\Sigma_{acdef}\Sigma_{bcdef}+\frac{\overline{\lambda}_3}{2\times
4!}
    S_{ab}\overline{\Sigma}_{acdef}\overline{\Sigma}_{bcdef}+\frac{\rho}{12}
    S_{ab}\Phi_{acde}\Phi_{bcde}
\eea

Note that ${\bf 54}= {\bf 15}_4 +\overline{\bf 15}_{-4}+{\bf
24}_0$ under SU(5). Therefore when $\sigma_o=\overline{\sigma}_o =
v_{B-L}$, the {\bf 15} multiplets have a $2\times 2$ mass matrix
of the form: \beq\label{15mass}
    \left(%
\begin{array}{cc}
  \frac{4}{5}m_{\Sigma} & \sqrt{2}\lambda_3\sigma_o \\
  \sqrt{2}\overline{\lambda_3}\overline{\sigma}_o & m_{15} \\
\end{array}%
\right)
\eeq
Similarly the {\bf 24} in {\bf 210} and in {\bf 54} mix and we have the
following mass matrix for the {\bf 24} Higgses:
\beq
    \left(%
\begin{array}{cc}
  m_{\Phi}+2 \lambda \widetilde{\Phi}_o & \sqrt{\frac{6}{5}}\rho\Phi_o \\
  \sqrt{\frac{6}{5}}\rho\Phi_o & m_{15} \\
\end{array}%
\right) \eeq There is no effect on the {\bf 45} and {\bf 50} Higgs
masses. We can now fine tune the {\bf 15} mass matrix to get one
{\bf 15}+$\overline{\bf 15}$ lower mass ( at $10^{13}$ GeV), while
keeping the other pair at the SO(10) scale. We could not have done
this without the {\bf 54} field. Furthermore, we fine tune the
parameters in {\bf 24} mass matrix to keep one {\bf 24} at the
SU(5) scale. Since the parameters in the {\bf 24} and {\bf 15}
mass matrices are different, the two fine tunings can be done
independently.

We thus conclude that in the minimal SO(10) model for the triplet term
to dominate type II seesaw formula, the minimal Higgs set required are:
{\bf 10}, {\bf 126}-pair, {\bf 210} and {\bf 54} dimensional.

We believe this result is interesting with important implications for
SO(10) model building.

\section{ NEW MINIMAL MODEL}

As already noted, the numerical analysis of neutrino sector indicates that
$f_{33}\sim 10^{-3}$ if the mixing angle $\alpha^u_2$ is of  order
1 and the mass of the triplet Higgs field
has to be order of $10^{12}$ GeV in order for the triplet term in type II
seesaw to
dominate. However, if the mixing angle can be reduced, $f$ will
increase and the the mass of the triplet can be higher. For
example, if the mixing angle is reduced by factor of  $10^{-4}$,
$f_{33}$ will increase by the same factor and the triplet mass
just needs to be $10^{16}$ GeV. While no fine tuning in Higgs {\bf 15}
mass is needed in this case, one needs a fine tuning in the values of the
Higgs mixing parameter that will allow an increase in the value of
$f_{33}$.  This is due to the fact that the fit that requires $r_1\sim
0.014$, $r_2
\sim 0.15$ and $\tan\beta = 10$, we need $\alpha^d_2 \sim
\alpha^u_2 \sim 0.0001$ and $\alpha^d_1 \sim 0.1$ in order to get
the required triplet mass up to GUT scale.

 Furthermore, making the triplet
of $\textbf{15}$ light by  fine tuning always leaves other
components heavy. In this case the coupling unification is
destroyed.

 Here we show in the model with $\textbf{54}$ Higgs the relative
lightness of {\bf 15} is achieved without bringing down any other
unwanted multiplet below the SO(10) scale. The symmetry breaking
scheme is two step type as before. First, SO(10) is broken at the scale of
$10^{18}$ GeV
down to SU(5) and then, to Standard model at the GUT scale
($2\times10^{16}$GeV).

\subsection{SUPERSYMMETRIC VACUUM}

The equations of $F_i=0$ with non-vanishing $v_{B-L}$ are

\bea
    m_{\Phi}S_-+6\lambda S_o^2+\eta \sigma_o\overline{\sigma}_o +2\rho SS_-=
    0\\\nonumber
    m_{\Phi}S_++2\lambda (S_+^2+2S_o^2)+\eta
\sigma_o\overline{\sigma}_o-\frac{4}{3}\rho SS_+ =
    0\\\nonumber
    m_{\Phi}S_o+2\lambda (S_-+2S_+)S_o+\eta \sigma_o\overline{\sigma}_o+
\frac{1}{3}\rho SS_o=
    0\\\nonumber
    m_{\Sigma}+\eta(S_-+3S_++6S_o)=0\\\nonumber
    \frac{5}{6}m_{15}S+\frac{5}{36}\lambda_1S^2+\rho(S_o^2+S_-^2-2S_+^2)=0
\eea where $S$ is the vev of the {\bf 24} in $S$ and $S_{\pm,o}$
are the linear combinations of the vev of {\bf 24}, {\bf 75} and
{\bf 1} in $\Phi$. They can be written explicitly as

\bea
    S_+&=& \frac{1}{3}<{\bf 75}>-\frac{2}{9}<{\bf 24}>+<{\bf 1}>\\\nonumber
    S_-&=& <{\bf 75}>+\frac{1}{3}<{\bf 24}>+<{\bf 1}>\\\nonumber
    S_o&=& -\frac{1}{3}<{\bf 75}>+\frac{1}{18}<{\bf 24}>+<{\bf 1}>
\eea

Note that $<\bf 1> = \widetilde{\Phi}_o$ breaks the SO(10) down to
SU(5) and other components break the SU(5) down to the MSSM. In
order for the two-step symmetry breaking to happen, we need to
have $S_i=\widetilde{\Phi}_o + m_{\Phi}s_i$ and
$S=\frac{m_{\Phi}}{\rho x}s$ ($x=\frac{\tilde{\Phi}_o}{m_\Phi}$)
with all $s_i$ and $s$ much less than 1 (they are of order
$10^{-2}$ in our scheme). The equations up to leading order become

\bea\label{lineareq}
     x+6\lambda x^2 +\eta
\frac{\sigma_o\overline{\sigma}_o}{m_{\Phi}^2}+s_-+12\lambda s_o x
+2 s=
    0\\\nonumber
    s_+-s_-+4\lambda x(s_+-s_o)  -\frac{10}{3}s=
    0\\\nonumber
    s_o-s_-+2\lambda x(s_-+2s_+-3s_o)  -\frac{5}{3}s=
    0\\\nonumber
    s_-+3s_++6s_o=
    0\\\nonumber
    \frac{5m}{6} s+2(S_o+S_--2S_+)=0
\eea where $m=\frac{m_{15}}{m_{\Phi}}\frac{1}{(\rho x)^2}$.

The first equation in (\ref{lineareq}) can be satisfied by solving
$\sigma_o\overline{\sigma}_o$. In order to have non-vanishing
solution of $s$'s, we require the condition

\beq
    Det\left(%
\begin{array}{cccc}
  1+4\lambda x & -4\lambda x & -1 & -4 \\
  4\lambda x & 1-6\lambda x & -1+2\lambda x & -2 \\
  3 & 6 & 1 & 0 \\
  -4 & 2 & 2 & m \\
\end{array}%
\right)=0 \eeq This gives $x\lambda=\frac{1}{4}$ or
$m=\frac{12}{1+2\lambda x}$. In the first case, the $\textbf{75}$
Higgs will be light and has huge contribution to the RG running
above the SU(5) scale. This will bring the coupling constant to
the strong regime below the required $10^{18}$ GeV. We will
exclude this and set $x\lambda \neq {-\frac{1}{2},\frac{1}{4}} $
from now on. The solution is given by

\beq
    \left(%
\begin{array}{c}
  s_+ \\
  s_o \\
  s_- \\
\end{array}%
\right)=\frac{s}{3(1+2\lambda x)}\left(%
\begin{array}{c}
  4 \\
  -1 \\
  -6 \\
\end{array}%
\right) \eeq Compared with the explicit forms of $s_{\pm,o}$, this
solution corresponds to $<\bf 75>=0$ and
$\frac{<\Phi^{24}>}{<S^{24}>}=-\frac{6}{1+2\lambda x}$ as a result
of our choice to keep the {\bf 75} heavy.

\subsection{effective SU(5) superpotential}
At the SU(5) level, the effective superpotential that emerges from
the SO(10) theory can be written as:
\begin{eqnarray}
W_{SU(5)}&=&h_dT\bar{F}\bar{F}_5+f_dT\bar{F}\bar{F}_{45}~ +h_u
TTF_5 + f_u TT{F'}_{5}\\
&+&h_\nu FF\bar{F}_{15}+h_D F\bar{F}_5 N +~W_{Higgs}
\end{eqnarray}
where the Higgs part of the superpotential $W_{Higgs}$ is given by
\begin{eqnarray}
W_{Higgs}~=~\sum_a M_a \chi_a\overline{\chi}_a~+~W_{Tr}
\end{eqnarray}
with $\chi_a$ going over all the multiplets. We only give the
trilinear terms of the form $({\bf 24})({\bf 15})(\overline{\bf 15})$
that will contribute a mass terms to the ${\bf 15}$ when the ${\bf 24}$
get a vev. We found
\beq\label{ST15mass}
    W_{Tr}~=~-\sqrt{3}\eta{\Phi_a}^b\Sigma_{bc}\overline{\Sigma}^{ac}
    +\lambda_1{S_a}^b S_{bc}S^{ac}+....
\eeq

With appropriate fine tuning, we can reduce the nonminimal SU(5)
group down to the MSSM where the Higgs fields are linear
combinations of the standard model doublets in {\bf 5} and {\bf
45} fields.

\subsection{MASS OF \textbf{15}}

Below the SO(10) scale $M_{10}$, the mass of $\textbf{15}$ is
given in Eq. (\ref{15mass})

\beq
    M_{15}=\left(%
\begin{array}{cc}
  \frac{4}{5}m_{\Sigma} & \sqrt{2}\lambda_3\sigma_o\\
   \sqrt{2}\overline{\lambda}_3\overline{\sigma}_o & m_{15} \\
\end{array}%
\right) \eeq

In order to have one light $\textbf{15}$, we require
$\frac{4}{5}m_{\Sigma}m_{15}=2\lambda_3\overline{\lambda}_3\sigma_o\overline{\sigma}_o
$. The two {\bf 15} can then be written in term of the light and
heavy {\bf 15}, ${\bf 15}_{L}$ and ${\bf 15}_{H}$. In term of the
light field ${\bf 15}_L$, Eq.(\ref{ST15mass}) become

\beq
    W_{Tr}~=~(8m_{\Sigma}^2\lambda_1{S}^{24}-25\sqrt{3}
\eta\lambda_3\bar{\lambda_3}\sigma_o\bar{\sigma}_o{\Phi}^{24})
    \overline{\bf 15}_L {\bf 15}_L +\cdot\cdot\cdot
\eeq where $\cdot\cdot\cdot$ stand for terms involve heavy
particle. This will give a mass of order $10^{16}$ GeV to the
light {\bf 15} and will destabilize the whole multiplet at the
scale of $10^{14}$ GeV. To stabilize the {\bf 15}, we set this
term to zero and that require

\beq
8m_{\Sigma}^2\lambda_1<{S}^{24}>=25\sqrt{3}\eta\lambda_3\overline{\lambda_3}
\sigma_o\bar{\sigma}_o<{\Phi}^{24}> \eeq
The corrections of the
{\bf 15} mass thus come from the higher dimensional operators,
which in general have contribution of order $\frac{<24>^2}{M_{10}}
\sim 10^{14}$ GeV. The splitting between different multiplets in
{\bf 15} are at most of this order and the effect on the
unification of coupling is like the small threshold effect.

\subsection{MASS OF \textbf{5}}

In our scenario, {\bf 45} is heavy and so its contribution to the
physical Higgs doublet comes from the integrating out the
$\textbf{45}$. The mixing angle $\alpha^d_2$ is therefore small
($\sim10^{-2}$). We now analyze the physical Higgs mass from
$\textbf{5}$'s in the approximation of SU(5) symmetry. The mass
matrix is given in Eq. (\ref{5mass}). Again, the determinant of
the matrix has to be zero. That gives

\beq \frac{\lambda}{\eta}=
\frac{2\alpha\overline{\alpha}\sigma_o\overline{\sigma}_o}{\eta
m_{H}\widetilde{\Phi}_o
+\alpha\overline{\alpha}\widetilde{\Phi}_o^2} \eeq

The small mixing of $\overline{\Sigma}_5$
($\alpha^u_2\sim10^{-2}$) require \beq
    \frac{2\sqrt{6}\bar{\alpha}\eta\widetilde{\Phi}}{\alpha\overline{\alpha}
\widetilde{\Phi}_o-\eta
    m_{H}}~=~100
\eeq

and $\alpha^d_1\sim 0.1$ require

\beq
    \frac{\alpha}{\overline{\alpha}}=10^3
\eeq

or

\beq
    \frac{\bar{\alpha}\widetilde{\Phi}_o}{\eta\sigma_o} ~=~ -10
\eeq

To summarize, we collect all conditions from the arguments above.

\bea
      \sigma_o\overline{\sigma}_o &=&-(x+6\lambda x^2+s_-+12\lambda s_o x+2
      s)\frac{m_{\phi}^2}{\eta}\\
      m&=&\frac{12}{1+2\lambda x}\\
      x\lambda &\neq& {-\frac{1}{2},\frac{1}{4}}\\
    \frac{4}{5}m_{\Sigma}m_{15}&=&2\lambda_3\overline{\lambda}_3\sigma_o\overline{\sigma}_o\\
    \frac{<\phi^{24}>}{<S^{24}>}&=&\frac{4\lambda_1m_{\Sigma}}{5\sqrt{3}\eta
    m_{15}}=-\frac{6}{1+2\lambda x}\\
    \frac{\lambda}{\eta}&=&\frac{2\alpha\overline{\alpha}\sigma_o\overline{\sigma}_o}{\eta
    m_{H}\widetilde{\phi}_o+\alpha\overline{\alpha}\widetilde{\phi}_o^2}\\
    \frac{2\sqrt{6}\bar{\alpha}\eta\widetilde{\phi}}{\alpha\overline{\alpha}\widetilde{\phi}_o-\eta
    m_{H}}&=&100\\
    \frac{\bar{\alpha}\widetilde{\phi}_o}{\eta\sigma_o} &=& -10
\eea

When these conditions are all satisfied, we have the required
triplet dominated type-II seesaw. Because $m$, $\lambda_1$ and
$\chi$ are free parameters, equation (42), (44) and (45) can be
satisfied by assigning the correct value to these three
parameters. Equation (47) can be satisfied by tuning the
denominator. We simply set the denominator of equation (47) to
zero and find that \bea
    \frac{\bar{\alpha}\widetilde{\phi}_o}{\eta\sigma_o} &=& -10 \\\nonumber
    \alpha\overline{\alpha}\widetilde{\phi}_o&=&\eta m_{H}\\\nonumber
    \frac{\lambda}{\eta}&=&\frac{\sigma_o\overline{\sigma}_o}{\widetilde{\phi}_o^2}=-\frac{1+6\lambda x}{\eta x}\\\nonumber
    \sigma_o\overline{\sigma}_o &=&-x(1+6\lambda x)\frac{m_{\phi}^2}{\eta}
\eea where we have simplified equation (41) by including only the
leading order terms. Note that
$x=\frac{\widetilde{\phi}_o}{m_{\phi}}$. We found from the
equations above that $\lambda x = -\frac{1}{7}$. If $m_H \sim
m_{\phi}$, we have also found the relations
$\frac{\alpha}{\overline{\alpha}}=\frac{7}{100}$ and
$\frac{\bar{\alpha}}{\sqrt{|\lambda \eta|}}=10$. There are enough
free parameters in the model to allow the above equations to be
satisfied simultaneously. In this model we have $f_{33}\sim 0.1$,
$v_{B-L}=10^{18}$ GeV, the triplet mass $M_T\sim10^{14}$ GeV and
the GUT scale remains at $2\times 10^{16}$ GeV.

One of the obvious concerns one may have with a {\bf 15}-Higgs
field around a mass of $10^{14}$ GeV is its contribution to proton
decay. As has been discussed very early on in the study of SU(5)
theories\cite{book}  {\bf 15} Higgs field cannot contribute to
proton decay in the limit of exact $SU(2)_L$. Therefore, typical
strength of proton decay amplitude arising from the exchange of
{\bf 15} Higgs is $\simeq \frac{v_{wk}h_\nu
h_d}{M_{15}M_5}\frac{m_{\tilde{G}}\alpha_s}{4\pi
m^2_{\tilde{q}}}\simeq h_\nu h_d 10^{-32}$ GeV$^{-2}$. This is far
below the present experimental limits on this strength.

\section{Gauge unification and SO(10) scale}

Given the above multiplet structure, we can now evaluate the
SO(10) unification scale. Since a viable type II theory requires
that SO(10) breaking be at least a factor of 5 bigger than the
SU(5) scale, we need to see if the gauge coupling remains
perturbative (i.e. $\alpha_U \leq 1$) until the SO(10) scale
$M_{10}$. For this purpose let us assume that the theory below the
SU(5) scale is MSSM as a starting point. Taking the contributions
of the various new Higgs fields above the SU(5) scale $M_5$, we
find that
\begin{eqnarray}
\alpha^{-1}_{10}~=~\alpha^{-1}_5 - \frac{b_5}{2\pi}ln\left(
\frac{M_{10}}{M_5}\right)
\end{eqnarray}
 where $b_5$ is the contribution  of the supermultiplets with masses
between the $SU(5)$ and the $SO(10)$ scale. In order to determine
$\alpha_{10}$, we need the value of $\alpha_5$ as well as the coefficient
$b_5$. The former will be different from the cannonical MSSM value since
now there is a full {\bf 15}-dim. multiplet below the SU(5) scale $M_5$.
Let us proceed to calculate this.

The running of the standard model gauge couplings from the
electroweak scale to the $SU(5)$ unification scale can be written
as
\begin{equation}
\alpha_i^{-1}(m_Z) = \alpha_5^{-1} +
\frac{b_i}{2\pi}\ln(\frac{M_5}{M_z}) + \frac{\delta_i}{2\pi}
\end{equation}
where $\alpha_{i= 1,2,3}$ are the properly normalized $U(1)$, $
SU(2)$ and $SU(3)$ gauge couplings, $\alpha_5$ and $M_5$ are the
$SU(5)$ gauge coupling and unification scale. The last term in
$(51)$ is the correction due to light multiplets (i.e with mass
smaller than $M_5$). In our model it corresponds to the $15 +
\overline{15}$ chiral superfields responsible for the type $II$
see-saw and is given by
\begin{equation}
\delta_i = \delta b_i\ln(\frac{M_5}{M_{15}})
\end{equation}
where
\begin{equation}
\overrightarrow{\delta b} = (7, 7, 7)
\end{equation}
Now it is straight forward to show that
\begin{equation}
\alpha_5 = \frac{\alpha^{(0)}_5}{1 + \alpha^{(0)}_5\Delta_5}
\end{equation}
$\alpha^{(0)}_5$ is the value of the $SU(5)$ gauge coupling with
the MSSM spectrum, and $\Delta_5$ is given by
\begin{equation}
\Delta_5  = -\frac{1}{56\pi}[5(\delta_2 - \delta_1) + 28\delta_2]
~=~-\frac{1}{2\pi}\delta_2
\end{equation}
With $M_{15} \sim 10^{14}\;GeV$ the gauge unification coupling
increases by $\sim 15\%$ which amounts to an increase of about
$30\%$ in the proton decay life time via the dimension $6$
operators. The $SU(5)$ unification scale and the value of
$\alpha_{strong}(m_Z)$ are unaffected by the extra light complete
multiplet. In Fig. 1, we show the coupling constant evolution in the case
with the {\bf 15} mass at $10^{13}$ GeV.

\bigskip

\begin{figure}[!hptb]
\begin{center}
 \includegraphics[height=3in,width=5in,angle=0]{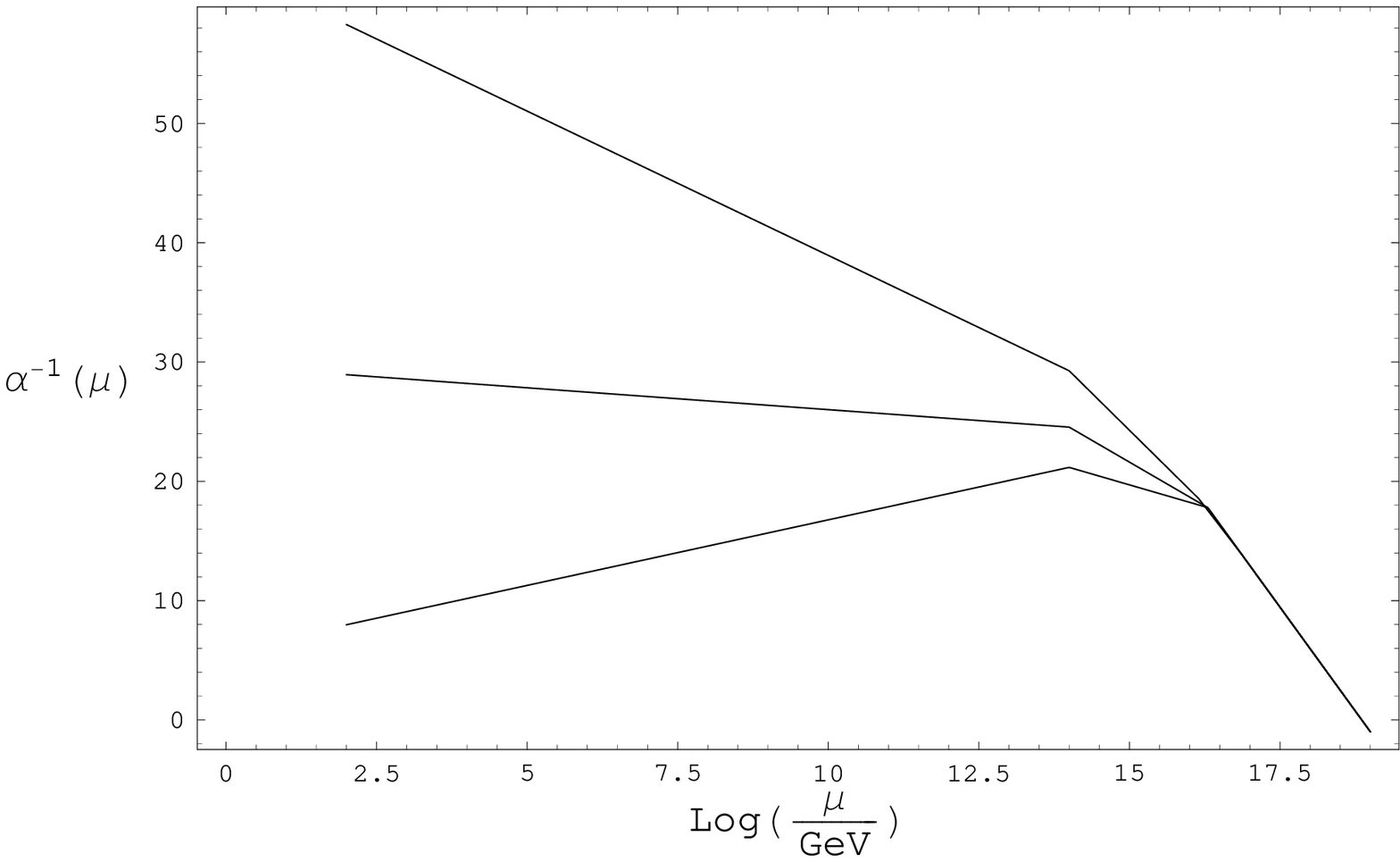}
\caption{\bf}
 \label{running}
  \end{center}
\end{figure}

\bigskip
\noindent{\bf Figure caption:} Running of the gauge couplings in the
presence of an SU(5) {\bf 15}-multiplet pair at $10^{14}$ GeV.

\bigskip

\begin{center}

{\bf Table II}

\bigskip

\begin{tabular}{|c||c||c|}\hline
$M_{15}(GeV)$ & $\alpha_5$ & $\tau^{(0)}_p/\tau_p $\\ \hline $10^{11}$ &
0.092 & 5.08 \\
\hline $10^{12}$ & 0.074 & 3.28
\\ \hline $10^{13}$ & 0.062 & 2.31 \\ \hline $10^{14}$ & 0.054 &
1.75
\\ \hline
\end{tabular}

\end{center}

{\bf Table Caption:}The $SU(5)$ gauge coupling and proton decay
with the $15-plet$ mass
$M_{15}=\{10^{11},10^{12},10^{13},10^{14}\}GeV$

\bigskip

Using this value of $\alpha_5$ we can evaluate $\alpha_{10}$ and check
whether it is perturbative at $M_{10}$. With one pair of $\{\bf 15, \bf
5\}$ and one ${\bf 24}$ we find $M_{10}\simeq 5.5\times 10^{18}$
 for $\alpha_{10}\simeq 1$. If we assume two ${\bf 24}$ and three pairs of
 ${\bf 5}$ below the $SO(10)$ scale
 we get $M_{10} \simeq 10^{18}$ GeV. Either of these values are sufficient
to make the second term in the seesaw formula small and make the triplet
term dominate the neutrino mass.



\section{CONCLUSION}
 We have given a thorough discussion of the
conditions under which the triplet term in type II seesaw
dominates the neutrino mass formula in a class of SO(10) models
with {\bf 126} that have been shown to lead to successful
predictions of neutrino masses and mixings.  This turns out to
impose nontrivial constraints on the nature of symmetry breaking
and Higgs structure of the model. For instance, we find that the
minimal Higgs structure consistent with requirements of gauge
coupling unification and triplet dominated type II seesaw is  a
combination of {\bf 210} and {\bf 54} multiplets in addition to
the multiplets {\bf 10} and the {\bf 126} pair required for
fermion masses. We give a detailed analysis of the Higgs
potential, the resulting mass pattern of Higgs fields as well as
the gauge coupling evolution in these models. We find that the
SO(10) must first break to a non-minimal SU(5) at a scale of about
$10^{18}$ GeV with SU(5) subsequently breaking down to MSSM at
$2\times 10^{16}$ GeV. A complete SU(5) {\bf 15}-plet must be
around a $10^{13}$ GeV scale to lead to both a triplet dominated
type II seesaw and gauge coupling unification. A low scale {\bf
15} does not lead to any new rapid proton decay modes. It can lead
to $\Delta B=2$ transitions such as neutron-anti-neutron oscillation
but the rate for this process in this model is too small to be
observable. We wish to emphasize that this symmetry breaking pattern
leaves our considerations for proton decay unchanged\cite{nasri}. It is
interesting that the triplet mass being of the order $10^{13}$ GeV
might generate a baryon asymmetry through its lepton violating
 interactions.

This work is supported by the National Science Foundation grant
no. Phy-0354401.


\begin{thebibliography}{99}

\bibitem{review}  A. Smirnov, hep-ph/0311259; S. F. King,
Rept.Prog.Phys. {\bf 67}, 107  (2004); G. Altarelli and F. Feruglio,
hep-ph/0405048; R. N. Mohapatra, hep-ph/0211252; NJP, {\bf 6}, 82 (2004).


\bibitem{so10} H. Georgi, in {\it Particles and Fields}
(ed. C. E. Carlson), A. I. P. (1975);
H. Fritzsch and P. Minkowski, Ann. of Physics, Ann. Phys. {\bf 93}, 193
(1975).


\bibitem{babu} K. S. Babu and R. N. Mohapatra, Phys. Rev. Lett. {\bf 70},
2845 (1993).

\bibitem{goran} B. Bajc, G. Senjanovi\'c and F. Vissani,
hep-ph/0210207; Phys. Rev. Lett. {\bf 90}, 051802 (2003).


\bibitem{goh} H. S. Goh, R. N. Mohapatra and S. P. Ng, hep-ph/0303055;
Phys.Lett. {\bf B570}, 215 (2003) and hep-ph/0308197; Phys. Rev. {\bf D 68},
115008 (2003).

\bibitem{mimura} For discussion of how to get CKM CP violation in these
models see B. Dutta, Y. Mimura and R. N. Mohapatra,
Phys.Rev.{\bf D 69}, 115014 (2004) and hep-ph/0406262 .

 \bibitem{seesaw2}  G. Lazarides, Q. Shafi and C. Wetterich,
Nucl.Phys.{\bf B181}, 287 (1981); R. N. Mohapatra and G. Senjanovi\'c,
Phys. Rev. {\bf D 23}, 165 (1981);

\bibitem{brahma} B.~Brahmachari and R.~N.~Mohapatra,
Phys.\ Rev.\ D {\bf 58}, 015001 (1998)
[hep-ph/9710371].
%

\bibitem{mimura1}  S.~Bertolini, M.~Frigerio and M.~Malinsky,
arXiv:hep-ph/0406117; Wei-Min Yang and  Zhi-Gang Wang, hep-ph/0406221.
 B. Dutta, Y. Mimura and R. N. Mohapatra, hep-ph/0406262.


\bibitem{minimal}  C. S.Aulakh and R. N. Mohapatra, Phys. Rev. {\bf D
28}, 217 (1983); T. E. Clark, T. Kuo and N. Nakagawa,
Phys. Lett. {\bf 115B}, 26 (1982); D. Chang, R. N. Mohapatra and
M. K. Parida,
Phys. Rev. {\bf D 30}, 1052 (1984); J. Baseq, S. Meljanac and
L. O'Raifeartaigh, Phys. Rev. {\bf D 39}, 3110 (1989); D. G. Lee,
Phys. Rev. {\bf D 49}, 1417 (1994); C. S. Aulakh, A. Melfo, B. Bajc,
G. Senjanovi\'c and F. Vissani, hep-ph/0306242.

\bibitem{recent} For discussion of these models for neutrinos, see
L.~Lavoura,
Phys.\ Rev.\ D {\bf 48}, 5440 (1993)
[hep-ph/9306297].
D.~G.~Lee and R.~N.~Mohapatra,
Phys.\ Lett.\ B {\bf 324}, 376 (1994)
[hep-ph/9310371];
D.~G.~Lee and R.~N.~Mohapatra, Phys.\ Rev.\ D {\bf 51}, 1353 (1995)
[arXiv:hep-ph/9406328];
K.~Matsuda, Y.~Koide, T.~Fukuyama and H.~Nishiura,
Phys.\ Rev.\ D {\bf 65}, 033008 (2002)
[Erratum-ibid.\ D {\bf 65}, 079904 (2002)]
[hep-ph/0108202];
T.~Fukuyama and N.~Okada,
JHEP {\bf 0211}, 011 (2002)
[hep-ph/0205066].
%

\bibitem{nath} P. Nath and R. Sayed, hep-ph/0103165; Nucl.Phys. B618
(2001) 138-156.

\bibitem{goran1} B. Bajc, A. Melfo, G. Senjanovic and F. Vissani,
hep-ph/0402122.

\bibitem{fukuyama} T. Fukuyama, A. Ilakovic, T. Kikuchi, S. Meljanac and
N. Okada, hep-ph/0405300.

\bibitem{aulakh} C. S. Aulakh and A. Giridhar, hep-ph/0405074.

\bibitem{book} See R. N. Mohapatra, {\it Unification and Supersymmetry},
3rd edition (Springer-Verlag, 2002), p. 125.

\bibitem{nasri} H.~S.~Goh, R.~N.~Mohapatra, S.~Nasri and S.~P.~Ng,
Phys.\ Lett.\ B {\bf 587}, 105 (2004)
[arXiv:hep-ph/0311330].

\end{thebibliography}
\end{document}